\documentclass[%
 aip,
 jmp,%
 amsmath,amssymb,
 reprint,%
]{revtex4-1}

\usepackage{graphicx}% Include figure files
\usepackage{dcolumn}% Align table columns on decimal point
\usepackage{bm}% bold math
\begin{document}

\preprint{AIP/123-QED}

\title{Reconfigurable Boolean Logic using Magnetic Single-Electron Transistors}% Force line breaks with \\
%\thanks{Footnote to title of article.}

\author{M. F. Gonzalez-Zalba}
 \email{mg507@cam.ac.uk}
 \affiliation{Hitachi Cambridge Laboratory, Cambridge CB3 0HE, United Kingdom}%
\author{C. Ciccarelli}%
	\email{cc538@cam.ac.uk}
\affiliation{Cavendsih Laboratory, University of Cambridge, Cambridge, United Kingdom}
\author{L.~P.~Zarbo}%
\affiliation{Institute of Physics, ASCR, v.v.i, Cukrovarnick'a 10, 162 53 Praha 6, Czech Republic.}
\author{A.C.~Irvine}%
\affiliation{Cavendsih Laboratory, University of Cambridge, Cambridge, UK}
\author{R.P~Campion}%
\affiliation{School of Physics and Astronomy, University of Nottingham, Nottingham NG7 2RD, United Kingdom}
\author{B.L.~Gallagher}%
\affiliation{School of Physics and Astronomy, University of Nottingham, Nottingham NG7 2RD, United Kingdom}
\author{T.~Jungwirth}%
\affiliation{Institute of Physics, ASCR, v.v.i, Cukrovarnick'a 10, 162 53 Praha 6, Czech Republic.}
\affiliation{School of Physics and Astronomy, University of Nottingham, Nottingham NG7 2RD, United Kingdom}
\author{A.J.~Ferguson}%
\affiliation{Cavendsih Laboratory, University of Cambridge, Cambridge, United Kingdom}
\author{J.~Wunderlich}%
\affiliation{Hitachi Cambridge Laboratory, Cambridge CB3 0HE, United Kingdom}
\affiliation{Institute of Physics, ASCR, v.v.i, Cukrovarnick'a 10, 162 53 Praha 6, Czech Republic.}
%\altaffiliation[Also at ]{Institute of Physics, ASCR, v.v.i, Cukrovarnick'a 10, 162 53 Praha 6, Czech Republic.}

\date{\today}% It is always \today, today,

\begin{abstract}
We propose a novel hybrid single-electron device for reprogrammable low-power logic operations, the magnetic single-electron transistor (MSET). The device consists of an aluminium single-electron transistors with a GaMnAs magnetic back-gate. Changing between different logic gate functions is realized by reorienting the magnetic moments of the magnetic layer which induce a voltage shift on the Coulomb blockade oscillations of the MSET. We show that we can arbitrarily reprogram the function of the device from an n-type SET for in-plane magnetization of the GaMnAs layer to p-type SET for out-of-plane magnetization orientation. Moreover, we demonstrate a set of reprogrammable Boolean gates and its logical complement at the single device level. Finally, we propose two sets of reconfigurable binary gates using combinations of two MSETs in a pull-down network. 
\end{abstract}

                             %% Classification Scheme.
                              %%display desired
\maketitle

\section*{Introduction}

As the downscaling of conventional CMOS technology is bound to reach its fundamental limit new algorithms will be the answer to achieve increasingly higher performance and reduced power consumption. Reconfigurable digital circuits provide a way to extend the functionalities of conventional CMOS by implementing in the same physical space multiple logic operations and therefore increasing the computational complexity. Reconfiguration of the logic functions at each individual device promises even more compact and flexible circuit design~\cite{Strukov2008,Joo2013,Heinzig2012,Yu2009,Mol2011}. However, the implementation of such reconfigurable logic using single-electron transistors (SETs)~\cite{Kastner1992, Gonzalez-Zalba2012} is appealing because SETs have good scalability, one of the lowest energy-per-switching-event~\cite{Zhu2007} and the possibility to combine their electrical properties with magnetic elements~\cite{Ono1997, Barnas2000, Shirakashi2001, Jalil2009, Takiguchi2014, Kirchner2009}.
There have been several proposals to implement programmable SET logic by using the charge degree of freedom such as fixed gate voltages \cite{Tucker1992}, non-volatile charge nodes \cite{Ishikuro1998, Uchida2002} and the spin degree of freedom \cite{Wunderlich2007, Dery2007, Hai2007}.

In this manuscript, we show a proof of principle for reconfigurable Boolean logic based on magnetically-gated circuit elements and we suggest multi-device reconfigurable logic architectures. More particularly, we report the complementary logic operation of a aluminium MSET. The reconfigurable capability of our MSET stems from the magnetization-dependent work function of GaMnAs back gate.  When the back-gate is kept at a constant potential the magnetic-field-induced chemical change causes a charge accumulation in the gate electrode which can be readily sensed by the SET as a change in the Coulomb oscillation phase~\cite{Wunderlich2006, Ciccarelli2012}. Although the concept is demonstrated here for MSET devices, the operating principle is general and transferable to any field-effect transistor.

\section*{Results}

The MSET has two modes of operation: it responds to gate voltage inputs (electric mode) as well as to the orientation of the magnetic moments (magnetic mode). By reorienting the magnetization of the GaMnAs substrate we are able to switch from n-type to p-type MSET. Making use of the magnetic mode we demonstrate two sets of reprogrammable Boolean logic gates implemented at the single device level. Finally, we suggest a strategy to scale the reconfigurable logic operation to the multiple device level.  

A schematic cross-section of the proposed aluminium MSET is depicted in Fig. 1(a). The key element of this structure is the magnetic gate electrode, an epitaxially grown Ga$_{0.94}$Mn$_{0.06}$As layer on GaAs that acts as a back-gate with easy-axis directions [110] and [1-10]. A magnetic field $B$=0.7~T, larger that the saturation field B$_{s}\approx$0.3~T, is applied to rotate the magnetization $M$ with respect to the [001] direction ($\phi$=0$^\circ$). The angle $\theta$ with respect to the [1-10] direction is kept equal to 90$^\circ$ in all measurements. The results presented here are independent of the magnitude of $B$ for $B>B_s$. A 100~nm alumina layer is used as gate dielectric on top of which the Al-SET is fabricated. The source and drain leads are separated from the micron-sized island by aluminium oxide tunnel junctions, as it can be observed in the SEM image in Fig. 1(b). In this letter the side gates were not used but could provide extra-functionality to the structure.

\begin{figure}[htbp]
	\centering
		\includegraphics{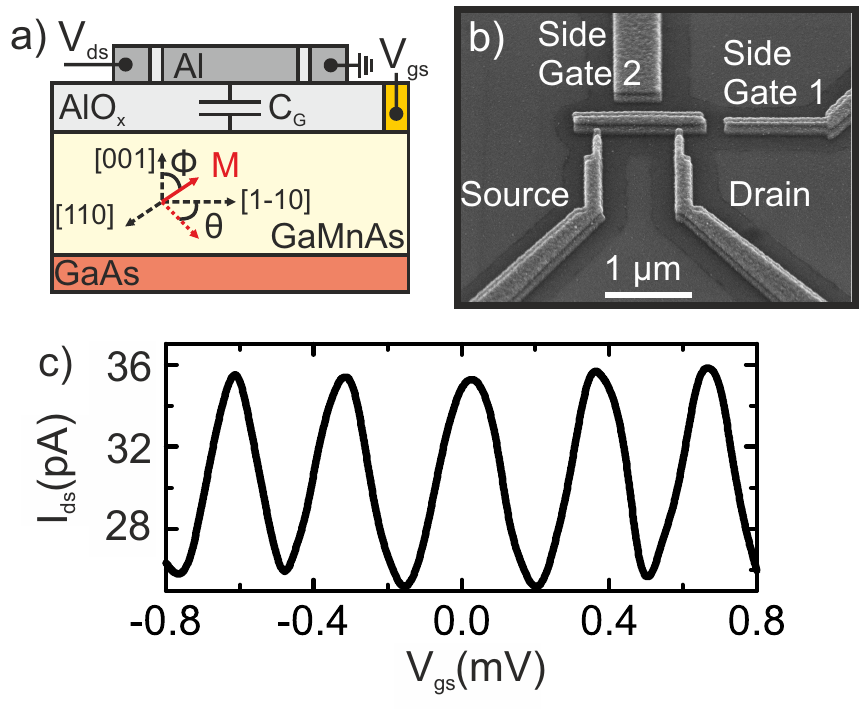}
	\caption{Device structure. (a) Schematic cross-section of the device sketching the magnetization orientation of the (Ga,Mn)As back-gate layer. (b) SEM image of the device. The aluminium island is separated from the source and drain leads by AlO$_{x}$ tunnel junctions. Side gates were not used in this experiment. (c) Drain current ($I_{ds}$) oscillations as a function of the back gate voltage ($V_{gs}$).}
	\label{fig:Fig1}
\end{figure}

In the electric mode of operation we use the back-gate voltage to control the induced-charge in the island of the MSET. Fig. 1(c) shows the drain current ($I_{ds}$) versus back-gate voltage ($V_{gs}$) at 300~mK. The plot shows a periodic oscillatory behaviour of the current, namely the Coulomb blockade oscillations. Several devices were studied with typical charging energies $E_{C}$=100-300~$\mu$eV and gate capacitances $C_{G}$=0.4-1.0~fF. We obtain a I$_{ON}$/I$_{OFF}$ contrast ratio of up to 1.33 and typical tunnel junction resistances of $R_T$=500~k$\Omega$.

The magnetic mode of operation is based on the dependence of the GaMnAs band structure on the orientation of the magnetization due to spin orbit coupling. The magnetization reorientation yields a shift in the GaMnAs work function $\Delta\mu$ which in itself does not produce a response on the MSET. However, when the back gate is held at constant potential a change in $\mu$ causes an inward or outward flow of charge in the gate which in turn offsets the Coulomb oscillations. If the work function shift in the magnetic layer is adequate, the phase shift becomes half a period of the Coulomb blockade oscillations and complementary MSETs are obtained. The equation that defines the necessary work function shift for maximum on-off current ratio can be expressed as,

\begin{equation}
	\Delta \mu (B)_{max}=\frac{e^2}{2C_G}
\end{equation}

As well, the minimum potential shift for complementary operation at temperature $T$ can be defined as a gate voltage shift equal to the FWHM of the Coulomb peak \cite{Beenakker1991},

\begin{equation}
	\Delta \mu (B)_{min}=\frac{4.35k_BTC_\Sigma}{C_G}
\end{equation} 

These two equations set the operational parameters in terms of temperature, magnetic field and physical dimensions of the structure.

The complementary operation of the MSET is experimentally demonstrated in Fig. 2 for a device with $E_C$=300$\mu$eV and $C_G$=1.0~fF. To switch between complimentary modes we rotate continuously a saturating magnetic field from $\phi$=90$^\circ$, where the magnetization resides $M$ parallel to the plane along the easy-axis [110] (Fig.2(a)) to $\phi$=0$^\circ$ where the magnetization is out-of-plane (Fig. 2(b)). This can be seen in Fig.2(c) where we plot the source-drain current $I_{ds}$ as a function of the magnetization angle $\phi$ and gate voltage $V_{gs}$ for $V_{ds}$=15$\mu$V and B=0.7~T. When the magnetization is in-plane, we select $V_{gs}$=-61~$\mu$V as the logic 0 and $V_{gs}$=+86~$\mu$V as the logic 1 for the n-type SET (see Fig. 2(d)). As the magnetic field is rotated to $\phi$=0$^\circ$, the magnetization is re-oriented in the [001] direction and the GaMnAs work function increases producing a gate voltage shift of $\Delta V_{gs}$=150~$\mu$V~\cite{workfunction}. The combination of work function potential shift in the Ga$_{0.94}$Mn$_{0.06}$As layer with the 300~$\mu$V gate voltage period of the device result in that the magnetically-induced gate voltage shift coincides with half a Coulomb oscillation (fulfilling eq.(1)). The MSET behaves effectively as a p-type SET, Fig. 2(e). Therefore we can arbitrarily select the function of the SET from an n-SET for in-plane magnetization $\phi$=90$^\circ$ to a p-type for out-of-plane magnetization $\phi$=0$^\circ$.

\begin{figure}[htbp]
	\centering
		\includegraphics{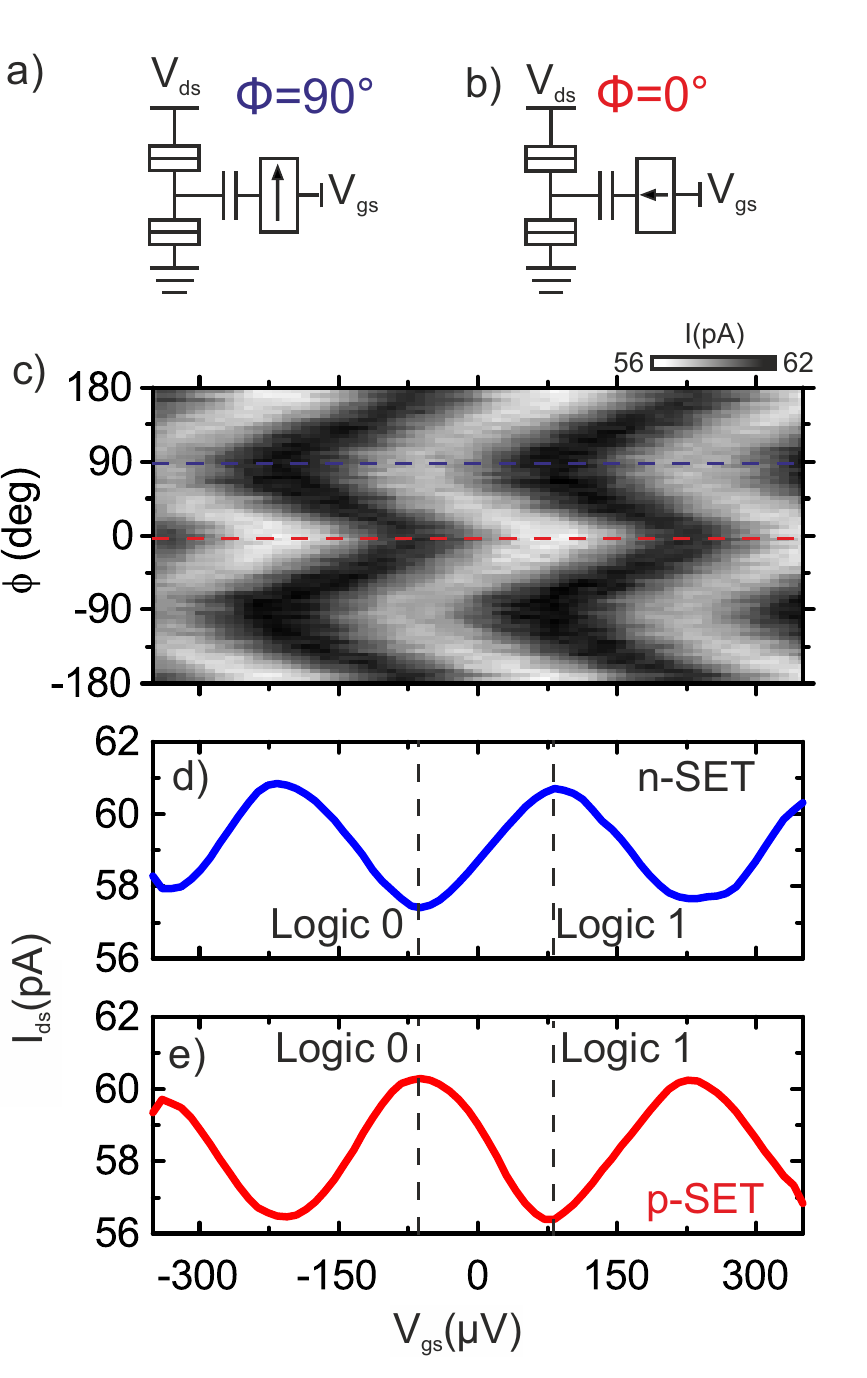}
	\caption{Complementary operation. Schematic diagram of the MSET for different magnetization orientations. (a) $\phi$=0$^\circ$ the magnetization is in-plane and (b) $\phi$=90$^\circ$ the magnetization is out-of-plane. (c) Coulomb blockade oscillations as a function of the direction of the back-gate voltage $V_{gs}$ and the applied magnetic field orientation $\phi$ for B=0.7~T. The-dashed blue and red lines indicate the operating points. (d) MSET Ids-Vgs transfer function at $\phi$=0$^\circ$. The logic 0 (1) has been selected at a low (high) current level, n-type SET. (e) MSET Ids-Vgs transfer function at $\phi$=90$^\circ$. The logic outputs have been inverted, p-type SET.}
	\label{fig:Fig2}
\end{figure}

\begin{figure}[htbp]
	\centering
		\includegraphics{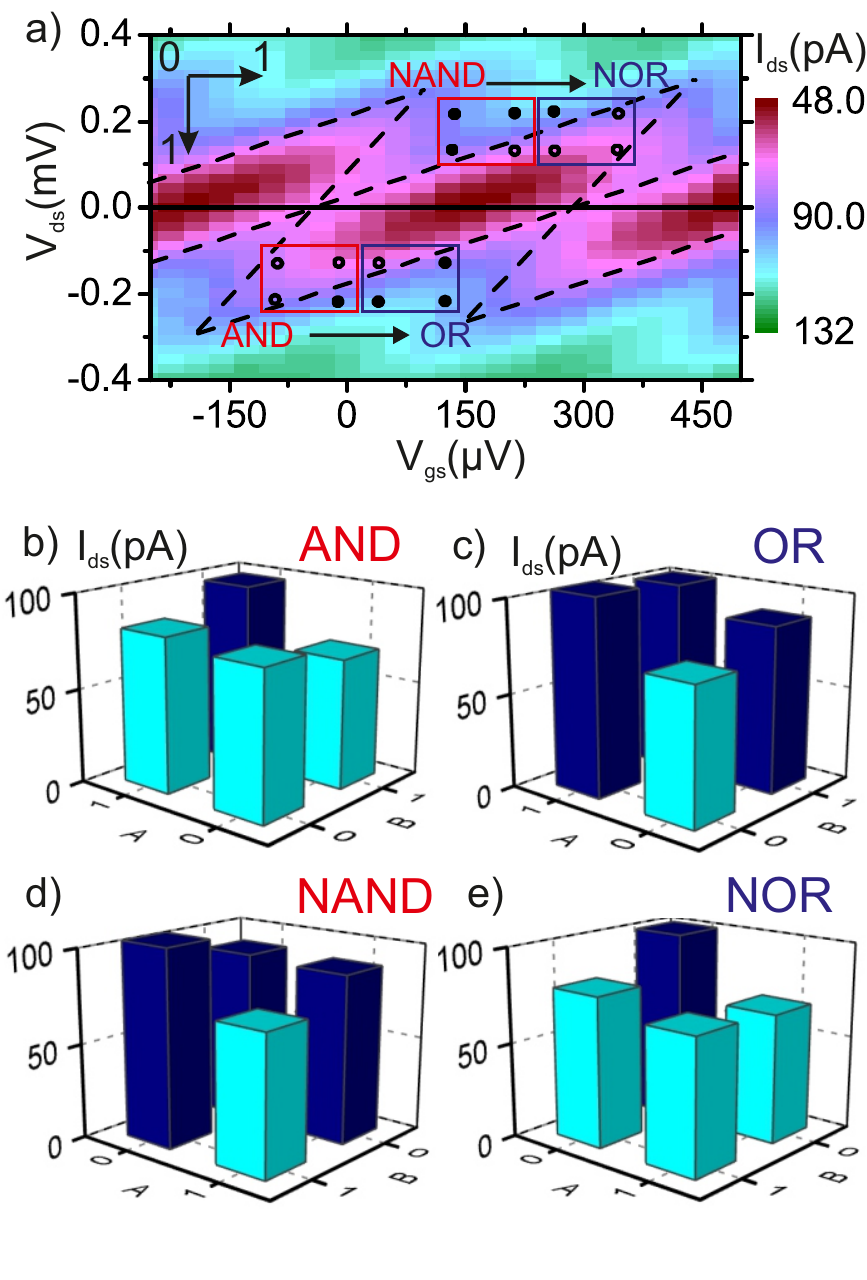}
	\caption{Single-device logic. (a) $V_{ds}-V_{gs}$  map of the drain current for $\phi$=0$^\circ$ showing the characteristic Coulomb diamonds. Red and blue frames sketch the implemented logic gates for $\phi$=0$^\circ$ and 90$^\circ$ respectively. (b-c) AND-OR set of reprogrammable logic gates. AND gate implemented at $\phi$=0$^\circ$ (b) and OR gate at $\phi$=90$^\circ$ (c) with $V_{ds}$ (input A) 0(1) defined as $-132(-220)$~$\mu$V and $V_{gs}$ (input B) 0(1) defined as $-96(0)$~$\mu$V. (d-e) NAND-NOR set of reprogrammable logic gates. NAND gate implemented at $\phi$=0$^\circ$ (d) and NOR gate at $\phi$=90$^\circ$ (e) with $V_{sd}$ (input A) 0(1) defined as 220(132)~$\mu$V and $V_{gs}$ (input B) 0(1) defined as 128(224)~$\mu$V.}
	\label{fig:Fig3}
\end{figure}

We now focus on the logic design that could be implemented at the single device level. It has been shown that any Boolean function of two variables can be implemented on a SET~\cite{Klein2009}. In this letter, by making use of the magnetic mode of operation, we demonstrate two sets of reprogrammable logic gates. The inputs in this case are the drain voltage (input A) and the gate voltage (input B) and the output is the drain current. In Fig. 3(a) we plot the $V_{ds}-V_{gs}$ diagram of the MSET showing the characteristic Coulomb diamonds at $\phi$=0$^\circ$. Overimposed, we schematically show two logic gates framed in red AND and NAND. The output is drawn as en empty dot if the results is 0 (low current level) and a full dot if the results is 1 (high current level). Upon changing the magnetization angle to $\phi$=90$^\circ$, the whole diagram is shifted horizontally by $\Delta V_{gs}$=-150~$\mu$V and, therefore, at the same input voltages the gate output changes to the two blue-framed logic gates, from (N)AND to (N)OR as depicted by the arrow. The reconfigurable logic gates are demonstrated in the histograms in Fig. 3(b-e). In order to discriminate between logic outputs 0 and 1 we select the low(high) current threshold at $I_{ds}$=80(90)~pA. In Fig. 3(b) we represent a histogram of the current output for an AND gate implemented on the edge of the Coulomb diamond. By rotating the magnetization to the in-plane direction ($\phi$=90$^\circ$) the Coulomb diamonds shift and the logic gate switches function to OR (Fig.~\ref{fig:Fig3}(c)). Moreover, in Fig.~\ref{fig:Fig3}(d,e) we demonstrate the logical complement set of reprogrammable Boolean gates. The set switches between NAND ($\phi$=0$^\circ$) and NOR ($\phi$=90$^\circ$) logic gates.

Finally, we briefly comment on the possibilities for multi-device reconfigurable logic design. As an example, we focus on pull-down networks formed by two elements with a common magnetic back-gate and independently addressable electrical gates. We concentrate on the case of MSET with identical $I_{ds}$-$V_{gs}$ transfer functions and logic inputs defined by Fig. 2(c,d). The circuit in Fig. 4(a) consists of a resistor load and two MSETs in series. In the state $\phi$=0$^\circ$ both MSET are in the p-type state and the network realizes the operation OR. However this gate can be reconfigured by rotating the magnetization, $\phi$=90$^\circ$. In this situation both MSET are in the n-type state and the logic operation becomes NAND. Similarly, for a pull-down network consisting of two MSETs in parallel, Fig. 4(b) the logic gate can be programmed to switch between AND ($\phi$=0$^\circ$) and NOR ($\phi$=90$^\circ$) operations. 

\begin{figure}
	\centering
		\includegraphics{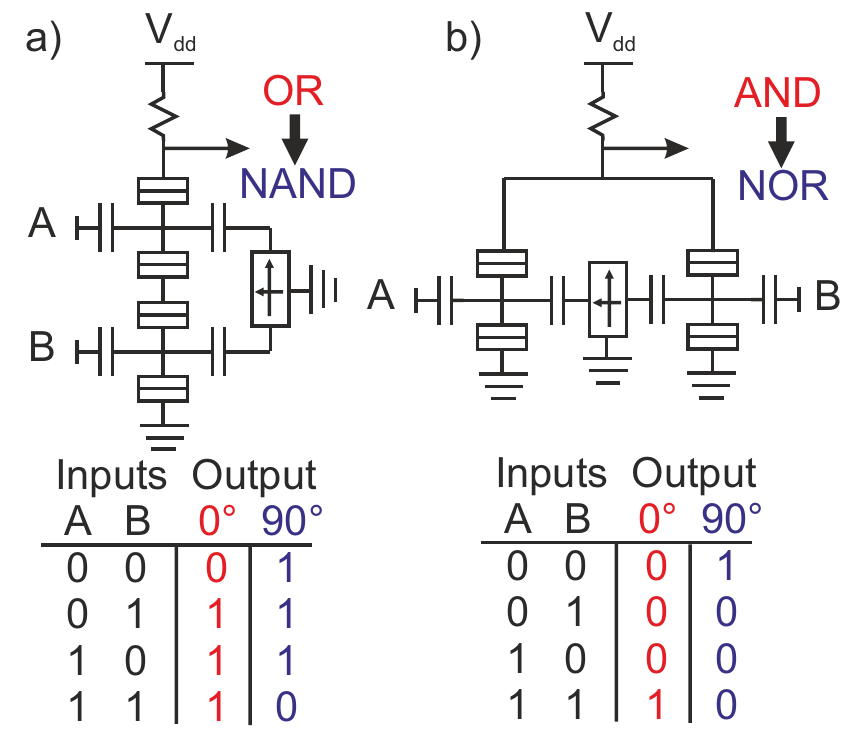}
	\caption{Logic at the multiple device level considering identical SETs and the logic inputs defined in Fig. 2. The inputs A and B are defined as taken as the SET gate values. (a) A series pull-down network performs the OR operation at $\phi$=0$^\circ$ and NAND at $\phi$=90$^\circ$. (b) Parallel pull-down network performs the AND operation at $\phi$=0$^\circ$ and NOR at $\phi$=90$^\circ$.}
	\label{fig:Fig4}
\end{figure}

\section*{Discussion}

In conclusion, we have demonstrated the complementary logic operation of an aluminium MSET by making use of magnetization-dependent work function of the GaMnAs magnetic back-gate. By using the source-drain voltage and the gate voltage as binary inputs we showed a set of Boolean gates and its logical complement implemented at the single device level, AND-OR and NAND-NOR gates. Finally, we have described a route to integrate multiple MSET in to integrated reconfigurable logic circuits. Although here we present results on MSETs, magnetic gating could be exploited to switch function in more conventional circuit elements such as CMOS field-effect transistors or novel electronic devices such as graphene ambipolar transistors~\cite{Wang2010} or spin-hot carrier transistors~\cite{Mizuno2007}. For reliable room temperature switching, magnetic materials with work function shifts of the order of 60~mV will be required~\cite{Betz2014}. Materials with large work function anisotropy such as CoPt alloys could provide that functionality~\cite{Wunderlich2006}. Moreover, non-volatile logic reconfiguration could be also achieved by making use of the stability of the magnetic moments along non collinear \lq\lq magnetic easy axes\rq\rq orientations without applying external magnetic fields or electrical currents.

\section*{Acknowledgements}
The authors thank M. Klein, D.A. Williams for fruitful discussion. The research leading to these results has been supported by the European Community's Seventh Framework under the Grant Agreement No.318397. http://www.tolop.eu, by the EU European Research Council (ERC) advance grant no. 268066, by the Ministry of Education of the Czech Republic grant no. LM2011026, and by the Grant Agency of the Czech Republic grant no. 14-37427G.

%\nolinenumbers
%
%%\section*{References}
%% Either type in your references using
%% \begin{thebibliography}{}
%% \bibitem{}
%% Text
%% \end{thebibliography}
%%
%% OR
%%
%% Compile your BiBTeX database using our plos2009.bst
%% style file and paste the contents of your .bbl file
%% here.
% 
%\begin{thebibliography}{10}
%\bibitem{bib1}
%Lorem M, Ipsum VE (1990) Rank Correlation Methods. New York: Oxford University Press, 5th edition.
%
%\bibitem{bib2}
%Ipsum M, Ipsum JD (1990) Rank Correlation Methods. New York: Oxford University Press, 5th edition.
%
%\end{thebibliography}


\begin{thebibliography}{26}
\bibitem{workfunction}
In-plane rotations between [110] and [1-10] directions were explored leading to $\Delta\mu$=46~$\mu$eV insufficient for complementary operation.
\input{workfunction}
\end{thebibliography}
\end{document}